\documentclass[12pt]{JHEP3}

\usepackage{mathrsfs}
\usepackage{amsmath,amssymb}
\usepackage{epsfig}
\input epsf

\def \no {\nonumber}

\def\l{{\lambda}}
 \def \const {{\rm const}}

\def\N{${\cal N}=4$ }

\def\be{\begin{equation}}
\def\ee{\end{equation}}

\newcommand{\bea}{\begin{eqnarray}}
\newcommand{\eea}{\end{eqnarray}}

\def\pa {\partial}

\def\la{\label}
\def\e{\epsilon}
\def\ov{\over}
\def\tr{{\rm tr}}

\def \bJ {{\rm J}}

\newcommand{\alg}[1]{\mathfrak{#1}}
\newcommand{\su}{\alg{su}}
\newcommand{\sls}{\alg{sl}}

\newcommand{\AdS}{{\rm  AdS}_5\times {\rm S}^5}

\newcommand{\sfrac}[2]{{\textstyle\frac{#1}{#2}}}

\def\ba{\begin{array}}
\def\ea{\end{array}}

\def \foot {\footnote}
\def \bi{\bibitem}
\def \tr {{\rm tr}}
\def \ha {{1 \over 2}}
\def   \td {\tilde}
\def \ci{\cite}
\def \N {{\mathcal N}}

\def \const {{\rm const}}

\def\ov{\over}

\def\l{\lambda}
\def\eps{\epsilon}

\def\foot{\footnote}

\def \l {\lambda}
\def\foot{\footnote}

\def \sql {{\sqrt \l}}

\newcommand{\rf}[1]{(\ref{#1})}
\def \ov {\over}

\def \ha{{1\ov 2}}
\def \sql{\sqrt{\l}}

\def \ep {\epsilon}

\preprint{ {\tt hep-th/0603113}\\ {\tt ITP-UU-06-09} \\ {\tt
SPIN-06-07}\\{\tt
Imperial-TP-AT-6-1}
}

\title{On highest-energy
 state in the $\su(1|1)$ sector of
 ${\cal N}=4$ super Yang-Mills theory}

\author{
G. Arutyunov$^{a}$\footnote{e-mail: G.Arutyunov@phys.uu.nl}
\footnote{Correspondent fellow at Steklov Mathematical Institute,
Moscow } and A. A. Tseytlin$^{b}$\footnote{Also
 at  The Ohio State University, Columbus  and
 Lebedev  Institute, Moscow.  }
\\
\\
$^{a}$ {\it Institute for Theoretical Physics and Spinoza Institute,
Utrecht University \\
~~3508 TD Utrecht, The Netherlands}\\
$^{b}$ {\it     Blackett Laboratory, Imperial College,
London SW7 2AZ, U.K.                      }\\
 }

\abstract{We consider  the highest-energy state in the $\su(1|1)$
sector of ${\cal N}=4$ super Yang-Mills theory containing
operators of the form tr$(Z^{L-M} \psi^M)$ where
$Z$ is a complex scalar and $\psi$ is a component of gaugino.
We show that this  state corresponds to the operator  tr$( \psi^L)$
and  can be viewed
as an analogue of the antiferromagnetic state in the $\su(2)$
sector.  We find perturbative expansions of the energy of this state
in  both weak and strong 't Hooft  coupling regimes
using asymptotic  gauge theory Bethe ansatz equations.
We also discuss  a  possible analog  of
this state in the  conjectured string  Bethe ansatz
equations.
 }

\begin{document}

\newpage

\renewcommand{\thefootnote}{\arabic{footnote}}
\setcounter{footnote}{0}
\section{Introduction}

Recent advances in understanding of  the AdS/CFT duality rely on
various kinds of evidence  that  both the large $N$ maximally
supersymmetric Yang-Mills theory \cite{mz1,bs} and the dual
classical $\AdS$ string theory \cite{Bena:2003wd,Arutyunov:2003uj}
are integrable models. An important part of the problem of
solving  the  conformal large $N$
 SYM  (or dual string theory)  would be  to compute the spectrum of
anomalous dimensions (or string energies) as explicit functions of
the 't Hooft coupling constant $\l$. These functions  should  describe a   smooth
interpolation from small $\l$ (perturbative gauge theory) region  to large
$\l$ (perturbative string theory) region.

Apart from a trivial case of  BPS operators whose conformal
dimensions are protected and, therefore, $\l$-independent, so far
we do not know any other operator for which the corresponding
dimension is exactly calculable. Only few    partial results   are
available.
  For  example, for the BMN-type  operators \cite{BMN} carrying the U(1)-charge $J$ under one of the
  U(1) subgroups of the internal symmetry group the string and gauge theory expressions
 for the anomalous dimension appear to coincide at leading order in the large $J$ expansion.
 Also, for low-energy  gauge spin chain states  dual to fast rotating strings  the
 two leading coefficients in the  large $J$, small $\l\ov J^2$  expansion
 happen to be the same on both sides of the duality  \ci{ft,kmmz}.
 In the  $\sls(2)$ sector,   few leading coefficients in  both small
 and large $\l$ expansions
 are known  for the  operators of the type $F D^S F$
 dual to the string spinning in ${\rm AdS}_5$, and
  one may  fit them approximately
  by  a simple  ``square root'' interpolating formula
  (using, e.g., the Pade
 approximation \ci{lip}), but this is  unlikely to be the
  exact answer.\foot{One may
  conjecture that it is
  more likely  to find a hypergeometric  function expression as in
 \ci{ryz}  where the dilatation operator was approximated by dropping all higher than
 spin-spin interaction terms.}

\medskip

 Recently, an interesting
 step towards finding the
 exact expressions for conformal dimensions was made:
 it  was pointed out  in \ci{rss,zar} that  it is possible to obtain a closed  expression for the
 energy of the highest  energy (``antiferromagnetic'') spin chain
 state in the $\su(2)$ sector by starting with the  asymptotic ``gauge theory''
  Bethe equations
  of BDS \ci{bds}  (that are supposed  to reproduce the
   gauge theory results up to  order $\lambda^L$ in the asymptotic
    expansion of the large spin chain length $L$).
 The resulting  expression for the  highest energy
 $\Delta (\l)$  found in the $L\to \infty $
 limit was given
 in terms of an
 integral of a  product of two  Bessel functions.
 It has a feature
 expected of strong-weak coupling   ``interpolating'' function:
  regular small $\l$ expansion is  smoothly
 connected to the $\sqrt \l$ asymptotics at large $\l$.
 Although this expression  need not
 match the exact  string theory
 expression,  one expects \ci{zar} to find a similar expression also from
 the genuine quantum string Bethe ansatz
 (which should presumably be of the AFS  ``string'' Bethe ansatz
 \ci{afs} type  modified  to  incorporate
   perturbative string results).
  In particular, the $\sql$  strong coupling asymptotics
  characteristic to the large  energy state
  is indeed
   found on the string theory side
 in the corresponding ``slow-string'' limit \ci{rtt}.

  \bigskip

Inspired by this  possibility of finding an exact expression  for the
conformal dimension in the $\su(2)$ sector one can try to extend
the work of \ci{rss,zar} by identifying similar special
 states in
other closed subsectors of the gauge theory.
In general, the spectrum of energies is unbounded in
non-compact sectors but there is   another
special choice  which is very similar to  the $\su(2)$ case:
 the
so-called $\su(1|1)$ sector
which is the simplest sector  containing   gauge-invariant
composite operators made of both bosonic and fermionic elementary
fields of ${\cal N}=4$ SYM theory.

\medskip

The goal of this paper is to identify
 an analogue of the  $\su(2)$ antiferromagnetic   state,
 i.e. the highest-energy state,
for the $\su(1|1)$ sector and compute  its conformal dimension
as a function of $\lambda$ both at weak and strong coupling
by starting again with the  asymptotic
gauge theory  Bethe ansatz equations of \ci{stau,beist}.

In contrast to the $\su(2)$  antiferromagnetic state for which
the form of the corresponding local operator  (tr$Z^{L/2} \Phi^{L/2} + ...$)
is  hard to describe  explicitly, here
the highest-dimension  operator is unique and
is easy to identify: it is the purely-fermionic one
tr$ (\psi^L)$. \foot{In the $\su(2)$
case the highest energy
antiferromagnetic  state  has complicated structure
and   can be effectively described only
by a density of distribution of roots in the large $L$ limit.
In contrast, the highest-energy state in the $\su(1|1)$ sector is
simple and the corresponding Bethe root distribution can be found
also for  finite $L$.}
At the same time,
 the  $\su(1|1)$  gauge  Bethe equations appear to be  more involved
than the ones describing the antiferromagnetic  state in the $\su(2)$ sector,
making the problem of finding a closed form of their  solution
rather non-trivial (and remaining unsolved  so far).
 Nevertheless, these equations appear to  admit   regular
perturbative expansions at  both small and large $\lambda$.


\medskip

An obvious next  question  concerns  realization of this
highest-energy state in string theory.
The conjectured ``string theory analog'' of the asymptotic
Bethe equations \ci{afs} is known to capture some  leading string
energy results  in certain
asymptotic expansions. One  is then tempted to
 try to find the
solution of these equations (which are similar
in both $\su(2)$ and $\su(1|1)$ sectors \ci{afs,beist})
which would correspond to a state with  highest possible energy.
In order to incorporate  quantum string corrections
\ci{bt,szz} the ``string  Bethe  equations'' of AFS  \ci{afs} should undergo
modifications beyond the
leading order and their
complete form
is currently  unknown.
Ignoring these modifications, one may still expect \ci{zar}
that the  AFS-type Bethe  equations  should  predict the same qualitative behaviour
for the highest-energy state as do  the gauge theory BDS equations.
 One implication of this is
 that  one should  change
a standard  $p_k \sim  { 1 \ov \sqrt[4]{\lambda}}$ assumption
 (characteristic to the so-called short strings)
about the  large $\l$ scaling behavior
 of  momenta  of  elementary
excitations that describe this highest-energy state
at large $\lambda$: $p_k $ should be approaching constant values at large $\l$.
Indeed, the  gauge theory ansatz predicts that
the energy of this state should scale as
$\sqrt{\lambda}$, while for short strings one finds $\sqrt[4]{\lambda}$ scaling
law  \cite{afs}.
 However, as we shall see below,  developing a
 consistent strong-coupling expansion of the AFS  equations in this case
is not straightforward.

  \medskip

Let us mention also that the   question about  the highest-energy state illustrates
  the
impossibility of an isolated treatment  of classically-closed string
sectors in quantum theory due to non-commutativity of the truncation and
quantization procedures.
 The ``reduced'' $\su(1|1)$ sector  of string theory
was described  in \cite{Alday:2005jm} as a consistent truncation
of the classical $\AdS$ superstring equations of motion and the
corresponding  quantum spectrum was then  found \cite{af2} by
quantizing this model in the light-cone gauge where it  becomes
equivalent to a theory of free fermions.
 Its  spectrum was  shown to contain both short and
long (winding) strings whose energies  scale as $\sqrt[4]{\lambda}$
and $\sqrt{\lambda}$ respectively; there is no apparent
  bound on the  energy
since the  energy of long strings can be
arbitrarily increased by increasing the winding number $m$.
What should presumably happen in the  full quantum superstring treatment is that
the string spectrum will become   periodic  in quantum numbers, so that
the states with  $m > L$   will be  equivalent to states with $m < L$
(the same should apply also to the $\su(2)$ sector case   \ci{rtt}).

\bigskip

The rest of this paper is organized as follows.
In  section 2 we will
identify an operator in the $\su(1|1)$ gauge theory sector which
 corresponds to
the highest-energy state of the  gauge theory  spin chain
and discuss perturbative solutions of the
 $\su(1|1)$  BDS-type  Bethe equations both at weak and strong
coupling.
In section 3 we shall comment  on  the search for a similar highest-energy
state  in the  ``string''  AFS-type Bethe ansatz equations
and  discuss  the conditions on
the scaling behavior of their solution which would lead to a
qualitative agreement with the gauge theory Bethe ansatz results.
Finally,  section 4 will contain a summary.

\renewcommand{\thefootnote}{\arabic{footnote}}
\setcounter{footnote}{0}
\section{Highest-energy state from asymptotic gauge theory Bethe ansatz}

The  simplest $\N=4$ SYM
 sector closed to all orders \ci{beis}
 is  the
$\su(1|1)$ sector. It contains operators of the form \be \la{ont}
{\rm tr}( Z^{L-M} \psi^M)  \ee with canonical dimension $
\Delta_0= L + { 1\ov 2} M $, the U(1)-charge $J= L - \ha M $ and
the Lorentz spin $ S=\ha M
 $ ($Z$ is a complex scalar and $\psi$ is the highest-weight component of the Weyl spinor
 from the vector multiplet). The integer $L$ is identified with
 the length of the corresponding spin chain  \ci{beii}.

The state we will be interested in has $M =L$, i.e.
corresponds to the operator
 \be \la{psy}
  {\rm tr} (  \psi^L ) \ , \ \ \ \ \ \ \ \ \ \
 \Delta=  { 3 \ov 2 } L +  {\cal O}(\l) \  . \ee
 This operator does not mix with other operators
  containing $Z$ and thus   should be  an eigenstate  of the
  dilatation operator.\foot{Indeed, the dilatation operator is built out of
    products of (super)permutation operators, so $\tr \psi^L$
    is an eigenstate just  like   $\tr Z^{L}$.
    The difference with the latter is the corresponding
    sign of  the
    eigenvalue  of the permutation operator  (which was +1 in  $\su(2)$
    case but is  -1 in $\su(1|1)$ case)  and as a result ${\rm tr} \psi^L$
    gets nontrivial anomalous dimension we aim to compute.}
  This state  provides a  simple example
   allowing to  check the consistency of solution of
  the  analog of the BDS Bethe ansatz for
   the $\su(1|1)$ sector  suggested  in \ci{stau,beist}.

Let us point out the following difference with the $\su(2)$ case.
There one studies the operators of the type ${\rm
tr}(Z^{J_1}\Phi^{J_2})$, where $\Phi$ is another complex scalar of
the ${\cal N}=4$ SYM and the length is $L=J_1+J_2$. The  highest-energy antiferromagnetic
 state
has $J_1=J_2$. There are {\it many} operators with the same
charges $J_1=J_2$ and the AF state is distinguished among them by
the requirement to have the maximal energy.
In the $\su(1|1)$ case
the state ${\tr}\psi^L$ is unique for  $ J= S= \ha L$, i.e. $M=L,$
 and it maximizes the energy on the  space of all  $\su(1|1)$
   operators with fixed $L$.

\subsection{Finite $L$ weak coupling results}

Our starting  point will be the all-order asymptotic
 Bethe ansatz for the
$\su(1|1)$ sector which is the  analog of the $\su(2)$ BDS Bethe
ansatz  \ci{stau,beist}
 \bea \la{ook}
 e^{ip_k L}=\prod_{k\neq
j}^{M}\frac{1-\frac{g^2}{2x_k^+x_j^-}}{1-\frac{g^2}{2x_k^-x_j^+}} \ , \ \ \ \ \ \ \ \ \ \
\ \ \ \   g^2 = { \l \ov 8 \pi^2}  \ . \eea
Here $M$ is the number of impurities, i.e. the number of $\psi$
operators in  \rf{ont}.
 The different quantities
entering eq.(\ref{ook}) are defined as \be
p_k=\frac{1}{i}\log\frac{x_k^+}{x_k^-} \ , \ \ \ \ \ \ \ \ \ \ \ \
\      x_k^\pm = x^\pm (u_k) \ , \ee \bea
x(u)=\frac{1}{2}\Big(u+\sqrt{u^2-2g^2} \Big) \ , \ \ \ \ \ \ \ \ \
~~~~~x^{\pm}=x\Big(u \pm \frac{i}{2}\Big)  \  . \la{ok} \eea Also,
\bea u(p)=\frac{1}{2}\cot{\frac{p}{2}} \sqrt{1+ 8 g^2
\sin^2\sfrac{p}{2}} \ , \la{ooo}  \eea
 \bea x^{\pm}(p)=\frac{e^{\pm
\sfrac{i}{2}p}}{4\sin\sfrac{p}{2}}\Big(1+\sqrt{1+   8 g^2
\sin^2\sfrac{p}{2}}\Big) \ . \la{yu}
\eea
Taking the
logarithm   of  \rf{ook} we find the following
equation for $p_k$
\bea \la{pok}
p_k=2\pi \frac{n_k}{L} +\frac{1}{iL}\sum_{j\neq
k}^M\log
\frac{1-\frac{g^2}{2x_k^+x_j^-}}{1-\frac{g^2}{2x_k^-x_j^+}}\, \  ,
\eea
where $n_k$ are integers.
The resulting dimension or energy is
\be \la{enn}
\Delta = L +  \ha  M  + E \ ,\ \ \ \ \ \ \ \ \ \ \ \ \ E=  i g^2 \sum^M_{k=1} \Big[
 \frac{1}{x^+(u_k)}-\frac{1}{x^-(u_k)}\Big]
\ee
or
\be
E=   \sum^M_{k=1} \Big[ \sqrt{1+ 8 g^2  \sin^2\sfrac{p_k}{2}} -1\Big]\ .
\la{jy} \ee
The total world-sheet momentum
\be  P = \sum_{k=1}^M p_k  =  2 \pi m \la{www} \ee
should be quantized due to the cyclicity of the chain
(the shift operator $
U=\exp(i\sum_k p_k)$
must be equal to the identity),  i.e.
the  physical solution $\{p_k\}$  of \rf{pok}  must be  such that the
``winding''  $m$  should  be integer.
Note that as long as $m$ is integer, there exists an equivalent
distribution $\{p_k\}$ for which $m$, i.e. the total momentum,  vanishes
(as it is usually assumed when  comparing to gauge-theory operators).
Indeed, both the BA equation \rf{ook},\rf{ooo} and the energy \rf{jy} are
invariant under shifts of each
$p_k$ by $2\pi$, i.e.  under
$$ p_k \  \to \ p_k + 2 \pi m_k\, , $$
where $m_k$ are arbitrary integers  (this is equivalent of
shifting $n_k$ in \rf{pok} by $m_k  L$). Then $m \to m + \sum_k
m_k$ and thus can be made to  vanish by  shifting, e.g., just one mode number.

As discussed  in \ci{stau}, the equation \rf{pok}
can be solved order by order in perturbation theory  in $g^2$.
At leading 1-loop order one has simply
$p_k = { 2 \pi n_k \ov L}$,  $k=1,..., M$,   where all $n_k$
must be different because of  Fermi statistics.
We  may restrict the numbers
 $n_k$ to belong to a ``fundamental region'', e.g.,  $[1,L]$, or, to
 allow the possibility to
 choose  $m=0$, to  $[-{L-1\ov 2}, {L-1\ov 2}]$.

The first two leading terms in the energy are then found to be \ci{BEI,stau}
\bea
E= &&4 g^2 \sum^M_{k=1} \sin^2\sfrac{\pi  n_k}{L} - 8 g^4 \bigg[
\sum^M_{k=1} \sin^4\sfrac{\pi n_k}{L}  \no \\
 && - { 2 \ov L}  \sum^M_{k,j=1}
 \cos\sfrac{\pi n_k}{L}\  \sin^2\sfrac{\pi n_k}{L} \
 \sin \sfrac{\pi n_j}{L}  \ \sin \sfrac{\pi (n_k-n_j)}{L} \bigg]   + {\cal O}(g^6) \ .
  \la{sth}  \eea
From  the 1-loop  term in the energy
it is clear that it is maximized  when $M=L$  so that
$n_k$ take all possible
distinct values from the interval    $[1,L]$,
i.e.
\be \la{kol}
n_k=k\  ,\   \ \ \ \ \  \ \ \ \ \
 p_k = 2\pi \frac{k}{L} +  {\cal O}(g^2) \  , \ \ \ \ \ \ \ k=1, ..., L \ .   \ee
Thus the maximal-energy state  in the spectrum, i.e.
the direct analog of the ``antiferromagnetic'' state in the  $\su(2)$ case,
should correspond to the purely-fermionic operator
\rf{psy}.
This unique state $ {\rm tr} \psi^L $  is   ``as far as possible'' from
the  lowest energy BPS state tr $Z^L$. Using \rf{kol} in
\rf{enn},\rf{sth}  we find \be \frac{\Delta}{L}=
\frac{3}{2}+2g^2-4g^4 + {\cal O}(g^6) \ . \la{lead}  \ee
 As a check of \rf{sth},
 the same  expression \rf{lead}  is  found  by directly applying the 1-loop and 2-loop
$\su(1|1)$ dilatation
operator  in eqs. (2.18) and (3.1) in \ci{stau}
to the state $ {\rm tr} \psi^L$   (for any finite $L$).\foot{If we  represent $Z$ as spin-up
state and $\psi$ as spin-down state,  then  only the
first $\pm (1- \sigma^3)$ terms  in eqs. (2.18) and  (3.1) in \ci{stau}
will contribute when the dilatation operator is applied to the $(01)$ spin-up state
corresponding to tr$\psi^L$.}

Notice  that we can now assume that $L$ is an odd number: for even $L$
 the operator  ${\rm tr}\ \psi^L$ vanishes identically (due to the  clash between
the cyclicity  of the trace and the anticommutativity of the ${\rm
SU}(N)$ adjoint matrix elements $\psi_{AB}$).  Thus the winding
number \rf{www} for  the momentum distribution  \rf{kol}
 \be
m = \ha  (L +1)        \   \la{win}
\ee
is indeed integer, and thus this momentum distribution
is consistent with periodicity.
 The   distribution  \rf{kol}  is indeed
 equivalent upon $-2 \pi$ shifts of upper half of the momenta
 to the one with $n_k$ in  the $[-{L-1\ov 2} ,{L-1 \ov 2}]$ interval:
 \be
 p_k = 2\pi \frac{k}{L} +  {\cal O}(g^2) \  , \ \ \ \ \ \ \ \ \ \ \ \ \
 k= -{L-1\ov 2} ,   ...,0, ...,{L-1 \ov 2}   \ .
    \ee


By iterating the equation \rf{pok} with the initial condition \rf{kol}
it is straightforward to compute  few subleading terms in
$p_k$ (for finite $L$)\foot{Note that
    $ \sum^L_{k=1} p_k = \pi (L+1)$ does not depend on $g^2$.}
\bea
p_k =&& { 2 \pi n_k \ov L} - g^2 \sin { 2 \pi n_k \ov L} + { g^4 \ov 4}   ( 10
\sin { 2 \pi n_k \ov L} + \sin { 4 \pi n_k \ov L}) \no \\
&&-\  { g^6 \ov 12}   ( 99 \sin { 2 \pi n_k \ov L} + 18  \sin { 4 \pi n_k \ov L}  + 4  \sin { 6 \pi n_k \ov L}
    )  + {\cal O}(g^8)   \ . \  \la{pp} \eea
    We determined  several higher-order terms in this expansion but do not give them explicitly here.
Then  the energy of this special state with $n_k=k$
computed using \rf{jy}  is found to be
 \bea
\nonumber
\frac{\Delta}{L}&=&\frac{3}{2}+2g^2-4g^4+\frac{29}{2}g^6-\frac{259}{4}g^8+\frac{1307}{4}g^{10}-
1790g^{12}+10396g^{14}\\
&&~~-\frac{504397}{8}g^{16}+\frac{6324557}{16}g^{18}-\frac{40702709}{16}g^{20}+
\ldots \ .
\la{eee}
 \eea
The growth of    the  coefficients
 is  an artifact of  expressing $\Delta/L$ in terms of $g^2 ={ \l \ov  8 \pi^2}$:
the coefficients  become numerically small when $\Delta$ is
expressed in terms of $\l$:
\bea
\frac{\Delta}{L}&=&\frac{3}{2}+ { \l \ov  4 \pi^2} -{ \l^2  \ov  16 \pi^4}
+ { 29 \l^3 \ov  1024 \pi^6} -  { 259 \l^4 \ov 16384  \pi^8}  +
{ 1307 \l^5 \ov  131072 \pi^{10}} -
 ... \ .
\la{qee}
 \eea
Note that the series  is sign-alternate
which is  consistent with a finite radius of convergence.\foot{Inverting the sign
of $\l$ one finds that  the series behaves in the qualitatively same way
as $1-\sqrt{1- x}$ expanded  in $x$ and truncated at finite order.}
The most naive approximation to an
exact expression  that has finite radius of convergence and
reproduces the  two  leading  weak-coupling expansion  coefficients
 is very simple to guess:
\be \la{ade} {\Delta_{\rm fit}\ov L}= { 1 }  +  \ha \sqrt{ 1 + { \l
\ov  \pi^2} }  \ , \ee \be \left({\Delta_{\rm fit}\ov L}\right)_{\l
\to 0} = { 3 \ov 2} + 2 { \l \ov 8\pi^2} - 4 { \l^2 \ov
(8\pi^2)^2}  + {32 \ov 2}   { \l^3 \ov (8\pi^2)^3} - ...\ . \
\la{ess} \ee The $\l^3$ coefficient here  $32\ov 1024$ is very
close to
 $29\ov 1024$  in \rf{qee}.

\subsection{Large  $L$  limit and equation for the  Bethe root  density}

In a  natural attempt to try to solve  the equation \rf{pok} exactly
and thus find the closed expression for  the energy \rf{eee} let
us  follow the  treatment of the  AF state of the $\su(2)$ chain in \ci{rss,zar}
and consider the  $L \to \infty$ limit.
In this limit we can take the continuum  approximation of \rf{pok}
getting the  following integral equation for  the density of roots:
 \bea
\frac{dp}{du}=-2\pi
\rho(u)+\frac{1}{i}\int_{-\infty}^{\infty}dv \rho(v) \frac{\pa}{\pa
u}\log
\frac{1-\frac{g^2}{2x^+(u)x^-(v)}}{1-\frac{g^2}{2x^-(u)x^+(v)}}\,
\ . \la{ui}
\eea
Here  the density  of Bethe roots $\rho(u)$
is defined as in  \ci{zar}  ($\xi= {k\ov L} \in (0, 1)$)
\be  \la{poke}
\rho(u)=-\frac{d\xi}{du} \ , \ \ \ \ \ \ \ \ \ \ \ \ \ \ \ \
\int_{-\infty}^{\infty} du \ \rho(u) =1    \ ,
\ee
and also
\be
 p(u) = \frac{1}{i}\log\frac{x^+ (u)}{x^-(u)} \ , \ \ \ \ \ \ \
 \frac{dp}{du}=  \frac{1}{i}
 \Big[
  { 1 \ov \sqrt{ ( u+ i/2)^2 - 2 g^2 } }
  - { 1 \ov \sqrt{ ( u- i/2)^2 - 2 g^2 } } \Big]  \ .   \la{ij}
\ee
$p(u)$  changes from $2 \pi$ to 0 while $u$ changes from $- \infty$ to $+
\infty$.
The energy  shift $E$ in \rf{enn}      is then  given by
\bea \la{jo}
{E\ov L}=
i g^2 \int_{-\infty}^{\infty} du\  \rho(u) \Big[
\frac{1}{x^+(u)}-\frac{1}{x^-(u)}\Big]  \ .
\eea
For comparison, the  linear integral equation
for $\rho (u)$  one finds in the $\su(2)$ sector
by starting with the BDS ansatz  is \ci{rss,zar}
 \bea \la{suu}
\frac{dp}{du}=-2\pi
\rho(u) - 2 \int_{-\infty}^{\infty}dv\  \rho(v)
  { 1 \ov ( u-v)^2 + 1 }
\ .
\eea
Eq. \rf{suu} is  obviously simpler than   \rf{ui}
which has  less trivial kernel and thus is not readily solvable by the Fourier
transform  (or by using \ci{hult}
the simple  rule of convolution of the two kernels
$ K= { 1 \ov ( u-v)^2 + 1 } $ which gives a similar kernel with shifted parameters
and thus
allows to invert  the operator $ I + { 1 \ov \pi} K$).
The solution of \rf{suu} found in \ci{rss,zar} is
($\bJ_n$ are Bessel functions)
\be  \la{old}  \rho(u) =
 { 1 \ov 2 \pi}
\int_{-\infty}^{\infty}ds  \   e^{isu}\
{ \bJ_0 ({ \sqrt \l \ov \pi} s ) \ov 2 \cosh { s \ov 2}} \ ,
 \ee
\be \la{ssuu}
  {\Delta_{\su(2)}  \over L} =  1+\frac{\sqrt{\lambda }}{\pi }\int_{0}^{\infty }
 \frac{ds}{s}\,\,\frac{\bJ_0\left(\frac{\sqrt{\lambda }}{2\pi }s\right)
 \bJ_1\left(\frac{\sqrt{\lambda }}{2\pi }s\right)}{\,{\rm
 e}\,^{s}+1}\ . \ee
While  \rf{ui} is simply a  linear integral equation,
we did not find a  way to solve  it in a closed form.
The weak-coupling perturbation theory leads to
the  following expression for the
 density
 \bea
\rho=\sum_{k=0}^{\infty}g^{2k} \rho_k  =\frac{1}{\pi}\sum_{k=0}^{\infty} g^{2k}
\frac { \sum_{m=0}^{2k} a_{km}  u^{2m}}{(u^2+\frac{1}{4})^{2k+1}} \,
\la{den}
\eea
The normalization condition  $ \int_{-\infty}^{\infty} du\ \rho(u) =1 $
gives
 \bea
\sum_{m=0}^{2k} 2^{-2m}  \Gamma(2k-m+\sfrac{1}{2})\Gamma(m+\sfrac{1}{2})
\  {a_{km}} =0\ .\ \ \ \
\eea
Explicitly,
\bea \nonumber
\rho(u)&=&\frac{1}{2\pi}\frac{1}{u^2+\frac{1}{4}}
+g^2
\frac{-5+48u^2+16u^4}{ 32\pi( u^2+\frac{1}{4}   )^3}+\\
&&~~~~ +g^4\frac{-31+1060u^2-1520u^4-320u^6+512u^8}{  512 \pi(
u^2+\frac{1}{4}   )^5}+\ldots \ .  \la{sss} \eea
 It is interesting that  all the coefficients $a_{km}$
 here are integers, compared to transcendental coefficients in the $\su(2)$ case
 (suggesting that a closed  form of the solution may be simpler than \rf{old}).
 Substituting \rf{sss} (with proper number  of higher-order terms included)
   into the expression for the energy
  \rf{enn},\rf{jo}
 we find  the same result \rf{eee}  as obtained  for finite $L$.

\subsection{Strong coupling expansion }

Let us now try to solve the Bethe  equation \rf{pok} or its large
$L$ version \rf{ui} in the strong coupling limit of   $g\sim \l \gg 1$. It is
useful to switch to momentum representation $u \to p$ (see
\rf{ooo}). We shall
assume that $p_k$ admit the  following  expansion
 \bea \la{gghh}
p_k=p_k^{(0)}+\frac{p_k^{(1)}}{\sqrt{\lambda}}+... \eea
 At strong coupling \rf{yu} gives
 (we omit the label $^{(0)}$ on $p_k$ for notational simplicity)
 \be \la{stro}
x^{\pm}\to\sfrac{\sql}{4 \pi} \ e^{\pm \sfrac{i}{2}p} \ep (p) +
...\ , \ \ \ \ \ \ \ \ \ep (p) \equiv {\rm sign}(\sin{ p\ov 2}) \
, ~~~~\eps(0)=0\, .\ee Then the  strong coupling limit of the
Bethe equations \rf{ook}
 becomes
\bea e^{ip_kL}= \prod_{j\neq k}^M\frac{1-e^{-\sfrac{i}{2}(
p_k-p_j)}\ep (p_k)\ep (p_j)}{1-e^{\sfrac{i}{2}(p_k-p_j)}\ep
(p_k)\ep
 (p_j)} + {\cal O}( { 1 \ov \sql})  \ . \la{hg}
 \eea
For $M=L$=odd these equations can be solved  as follows.
Let assume that  there exists a solution with all $ \ep (p_k) =1$
(this does not restrict the generality:
if for some $p_k$  we have $ \ep (p_k) =-1$
we can shift it by $2 \pi$ and change this sign).
Then
\bea
e^{ip_kL}=
   \prod_{j\neq k}^L   e^{ i \pi   -\sfrac{i}{2}(p_k-p_j)  }  + {\cal O}( { 1 \ov \sql}) \ ,   \eea
and  thus (since   $ (e^{i \pi})^{L-1} =1$  for $L$ odd)
 \bea \frac{3}{2} p_k  L = 2 \pi  n_k  + \pi m  +  {\cal O}( { 1 \ov \sql}) \ , \ \ \ \ \ \ \ \
 m \equiv \frac{1}{2\pi }\sum_{j=1}^L p_j         \, .
\eea
From here $m= { 1 \ov L} \sum^L_{k=1}  n_k  $, and this number must be integer.
Assuming, as at weak coupling,
that   $n_k =k=1, ..., L$ we get $m= \ha (L+1)  $
which is indeed integer for  odd $L$.
As a result, we find a consistent solution for momentum distribution
at strong coupling
\bea
p_k=\frac{4\pi k }{3L}  + { \pi \ov 3}  \frac{L+1 }{L } \  ,~~~~~~~~k=1,\ldots , L \, . \la{soli}
\eea
Here all $p_k$ lie  on the interval $(0, 2\pi)$  so
 with this choice one has indeed sign$(\sin{p_k\ov 2})=1$.
One  can also choose an equivalent distribution with $m=0$ by shifting
one or few momenta by  multiples of $2 \pi$.
 For example, for $L=3$   a choice with $m=0$ is
 $p_k= (- { 20 \pi \ov 9}, { 8 \pi \ov 9}, { 12 \pi \ov 9})$.

 At the
subleading  $\frac{1}{\sqrt{\lambda}}$ order we find that the
logarithm of the Bethe equations takes the form
 \bea\la{solp}
 p_k^{(1)}=\sum^L_{j\neq
k}\Big[ \frac{p_j^{(1)}-p_k^{(1)}}{2L}+\frac{\pi}{L}\frac{\Big(
\frac{\eps(p_k)}{\sin\frac{p_k}{2}}+\frac{\eps(p_j)}{\sin\frac{p_j}{2}}
\Big)\sin\frac{p_j-p_k}{2}}{1-\cos\frac{p_j-p_k}{2}\eps(p_j)\eps(p_k)}
\Big] \ , \ \ \ \ \ \ \  p_k\equiv p_k^{(0)}
 \eea and it allows one to determine the correction to the leading
momenta   \rf{soli}. If we  assume that $\sum^L_k p_k^{(1)}=0$ then we obtain \bea
p_k^{(1)}=\frac{2\pi}{3L}\sum^L_{j\neq k} \frac{\Big(
\frac{\eps(p_k)}{\sin\frac{p_k}{2}}+\frac{\eps(p_j)}{\sin\frac{p_j}{2}}
\Big)\sin\frac{p_j-p_k}{2}}{1-\cos\frac{p_j-p_k}{2}\eps(p_j)\eps(p_k)}\,
. \eea Since all $\eps(p_k)=1$, then  finally
\bea
p_k^{(1)}&=&\frac{2\pi}{3L}\sum_{j\neq k}^{L}
\cot\frac{p_j-p_k}{4}\Big(
\frac{1}{\sin\frac{p_k}{2}}+\frac{1}{\sin\frac{p_j}{2}} \Big) \, ,
\eea
where indeed  $\sum^L_k p_k^{(1)}=0$.

The strong-coupling expansion of the energy \rf{jy} is
\bea
E=\frac{\sqrt{\lambda}}{\pi}\sum^L_{k=1}|\sin\frac{p_k}{2}| \ -L+\frac{1}{2\pi}\sum^L_{k=1}
p_k^{(1)}\cos\frac{p_k}{2} \ \eps(p_k)  \, +
 {\cal O}( { 1 \ov \sql})  \ . \la{hyg} \eea
In the present case we  find at large $L$
\be
{\Delta  \ov L}=  {E  \ov L}  +  { 3 \ov 2}  =
   c_1 \sqrt{\lambda}   +  c_2  +  {\cal O}( { 1 \ov \sql})  \ ,  \ \ \ \ \ \
\la{qqw}
\ee
\be (c_1)_{_{L \to \infty}} = \frac{3\sqrt{3}}{2\pi^2} \approx  0.26  \ ,  \la{ccc} \ee
\be
  (c_2)_{_{L \to \infty}}  = { 1 \ov 2}  +  \Big[
\frac{1}{3 L^2}\sum^L_{k,j=1}    \cot\frac{p_j-p_k}{4}\Big(
\frac{1}{\sin\frac{p_k}{2}}+\frac{1}{\sin\frac{p_j}{2}} \Big)  \cos\frac{p_k}{2}
\Big]_{_{_{L \to \infty}}}\approx   1.18
   \ ,
\la{cvv} \ee where to compute
 $ (c_2)_{_{L \to \infty}} $  we used
\rf{soli} and numerically evaluated the sum.

For comparison, in
the $\su(2)$ case  by starting from the   exact solution
at infinite $L$ and  weak coupling $\l$ \rf{ssuu} and interpolating to
 strong coupling  one finds
\be \la{sde} \left( {\Delta_{\su(2)}\ov L} \right)_{L \to \infty, \ \l \to \infty  }
 =  { \sql \ov  \pi^2 } +  { 3 \ov 4}  +  ...  \ ,
\ee where   dots  stand for  exponentially small
corrections.\foot{We thank A. Tirziu and K. Zarembo  for a
discussion of this expansion.} This seems to suggest that the
strong-coupling expansion of the solution of the BDS-type  Bethe
ansatz  may turn out to be only asymptotic also in other
sectors.\foot{We thank M. Staudacher for suggesting  this  to us.}
In the  absence of a closed   expression for the energy, i.e.  the
$\su(1|1)$ counterpart of \rf{ssuu} in the $\su(2)$ sector, we are
unable to decide if the expansion in \rf{qqw} will or will not
contain exponential corrections.\footnote{These  exponential
corrections are likely to be an artifact of the BDS ansatz related
to the order of limits problem and might be absent on the string
theory side where  there is no an apparent  reason for $e^{-{ 1
\ov \sql}}$-terms
 (no world-sheet instantons, etc). In general,
the   interpolation from weak coupling (gauge-perturbative region)
 to strong coupling  (string perturbative region)
 should be
done in  the full expression for the  energy $E(\lambda, L)$ which
need not be the same as the one coming out of  the asymptotic BDS
equations. } Still, it is amusing to note that the most naive
interpolation formula
\rf{ade}
that
reproduced exactly the first  two  leading  weak-coupling expansion  coefficients
gives also a relatively
 good fit at strong coupling:
\be \la{sade}
\left({\Delta_{\rm fit}\ov L}\right)_{L \to \infty, \ \l \to \infty  }
 =  { \sql \ov 2 \pi } +  1 +  {\cal O}( { 1 \ov \sql})
 = 0.16  \sql  + 1 +  {\cal O}( { 1 \ov \sql})  \, ,
\ee
where the coefficients $0.16$ and 1 are not that far from $ 0.26$ and $1.18$
in \rf{cvv}.

\bigskip

One may  question  if the above strong coupling solution for the
distribution of $p_k$ is unique.\foot{We thank  D. Serban for
stressing  this issue.} It may seem indeed that for finite $L$ one
may find many similar solutions with different momentum range.
However, one should remember that  the asymptotic BDS-type  ansatz
is related to gauge theory only to order $\l^L$, i.e. keeping $L$
finite while taking $\l$ large  may not be  consistent. One might
further expect that different  solutions which admit regular large
$L$ limit will, in fact, be equivalent, i.e. will lead {\it in
this limit} to the same expression for the energy. To illustrate
this point, let us mention  that there exists another solution of
the strong coupling Bethe equations (\ref{hg}) which has momenta
$p_k$ symmetrically distributed around zero \bea \nonumber
p_k&=&\frac{4\pi
k}{3L-1}+\pi\frac{L-1}{3L-1},\ ~~~~~~~~k=1,\ldots , (L-1)/2\\
\label{ns}
p_0&=&0\, ,\\
\nonumber p_k&=&\frac{4\pi
k}{3L-1}-\pi\frac{L-1}{3L-1}\ ,~~~~~~~~k=-(L-1)/2,\ldots , -1\, .
 \eea
For a  symmetric distribution the Bethe equations (\ref{hg}) take
the form \bea e^{ip_kL}=\prod_{j=1}^{\sfrac{L-1}{ 2 }}
\frac{1-e^{-\sfrac{i}{2}( p_k-p_j)}}{1-e^{\sfrac{i}{2}(p_k-p_j)}}
\frac{1+e^{-\sfrac{i}{2}(
p_k+p_j)}}{1+e^{\sfrac{i}{2}(p_k+p_j)}}\, , ~~~~k= 1,\ldots ,
(L-1)/2 \, , \eea which are indeed solved by (\ref{ns}). In the
large $L$ limit the momenta spread over the interval
$(-\pi,-\frac{\pi}{3})\cup (\frac{\pi}{3},\pi)$. The energy of
this solution in the large $L$ limit has the {\it same} leading
term, eq.(\ref{ccc}), the subleading terms are however different.
Without further input one can not decide if the solutions we found
indeed correspond to the highest-energy state as it is seen from
the strong coupling perspective.

 \bigskip

The  same  result
 for the leading strong-coupling term in the
energy \rf{qqw}   can be found also
from the strong-coupling limit of the
integral equation \rf{ui}   after converting it into
the ``momentum'' form:
\bea
1=2\pi   \td \rho(p)
+ \frac{1}{i}\int  dq \ \td \rho(q) \frac{\pa}{\pa
p }\log
\frac{1-\frac{\l}{(4 \pi)^2  x^+(p)x^-(q)}}{1-\frac{\l }{ (4 \pi)^2  x^-(p)x^+(q)}}\,
\ . \la{uir}
\eea
Here the momentum density is (recall that $ { du \ov dp } <  0$)
\be
 \td \rho(p)  \equiv - { du \ov dp } \rho (u) \  ,
\ \ \  \ \ \ \ \ \ \ \  \int^{p_{max}}_{p_{min}}  dp \ \td \rho(p) =1
 \la{dens} \ ,  \ee
and  the  limiting values $p_{max},p_{min}$ may,  in general,
depend on $\l$. Using \rf{stro} we get   (assuming that within
$(p_{max},p_{min})$ $\sin\frac{p}{2}$ has positive sign) \bea
1=2\pi \ \td \rho(p) + \frac{1}{i}\int^{q_{max}}_{q_{min}}    dq \
\td \rho(q) \frac{\pa}{\pa p  }\log \frac{1- e^{ - {i \ov 2}
(p-q)}
 }
{1-    e^{  {i \ov 2} (p-q) }
 }+  {\cal O}( { 1 \ov \sql}) \,
.  \la{irg} \eea From  here $ 1=2\pi \ \td \rho(p)  -  { 1\ov 2 }
+   {\cal O}( { 1 \ov \sql})$, i.e. \be \la{rrrr} \ \td \rho(p) =
{ 3 \ov 4 \pi } +   {\cal O}( { 1 \ov \sql}) \ .  \ee Thus the
momentum density is constant as it was at weak coupling ($\td
\rho(p) =  { 1 \ov 2 \pi }$, see \rf{pp}), but the momentum distribution range
have changed from $2 \pi$ to  $\frac{4\pi}{3}$. This is exactly
what  we have  found above in the discrete (finite $L$)
approach \rf{soli}. To match \rf{soli} we are to choose (at $L \to
\infty$)\  $p_{min}={ \pi \ov 3},\ p_{max}= { 5\pi \ov 3}
$.\foot{The momentum interval is fixed so  that its length
is $\frac{4\pi}{3}$ to have $ \td \rho(p)$ normalized and also to
satisfy the assumption that $\sin \frac{p}{2}$ has positive sign.}

Similar observations were made in the $\su(2)$ sector in \ci{zar},
where  the momentum quantization   condition and thus  the coefficient
 in the momentum
density had   changed by the factor of 2 in  going from the
weak to strong coupling.\foot{In the $\su(2)$ sector one can  get  (from   \rf{old})
 a closed  formula  for
$\td \rho (p)$  as a function of $\l$  that  interpolates between the  two
constant values at $\l=0$ and $\l= \infty$.}

The general expression for the energy
in the continuum limit in the momentum form is (see \rf{enn},\rf{jo})
\be
{\Delta\ov L}=  \ha  +
\int^{p_{max}}_{p_{min}}  dp \ \td \rho (p)  \
 \sqrt{ 1 + {\l \ov \pi^2} \sin^2 { p \ov 2} }   \ ,
 \la{coo}
\ee
so that at  strong coupling we get the same result as in \rf{qqw},\rf{ccc}
\be
\left({\Delta\ov L}\right)_{\l \to \infty }=
  {\sqrt \l \ov \pi} \int^{  { 5\pi \ov 3}  }_{ { \pi \ov 3}     }
   dp \ \td \rho (p)  \   | \sin { p \ov 2}  |
+ ... =    {3 \sqrt 3  \ov 2\pi^2  } {\sqrt{\lambda} } + ...    \,
,
 \la{coog}
\ee
where we used  \rf{rrrr}.

\section{Remarks on highest-energy state in the spectrum of
  the ``string'' Bethe ansatz}

In \cite{afs} a novel type of the  Bethe ansatz equations was
introduced to describe the leading quantum corrections to the
spectrum of classical strings on ${\rm AdS}_5\times {\rm S}^5$.
{\mbox Originally} conjectured for the $\su(2)$ sector, this
ansatz was subsequently generalized to other sectors of rank one
\cite{stau}, and finally to the full string sigma model on ${\rm
AdS}_5\times {\rm S}^5$ \cite{beist}.

If $s=-1,0,1$   for the $\sls(2)$, $\su(1|1)$  and  $\su(2)$ sectors
then the
 conjectured ``quantum string''  Bethe ansatz equations can be written
 in the form
\cite{beist}
\bea \la{ei}
 e^{ip_k L}=\prod_{k\neq
j}^{M}\
  \left(\frac{x_k^+ -  x_j^-}{x_k^-      -  x_j^+} \right)^s\
    \frac{1-\frac{g^2}{2x_k^+x_j^-}}{1-\frac{g^2}{2x_k^-x_j^+}} \  e^{ i \theta(p_j,p_k) }    \ ,
 \\
  e^{ i \theta(p_j,p_k) }=
   \left( \frac{1-\frac{g^2}{2x_k^+x_j^-}}{1-\frac{g^2}{2x_k^-x_j^+}}\right)^{-2}
   \  \left( \frac{1-\frac{g^2}{2x_k^-x_j^+}}{1-\frac{g^2}{2x_k^+x_j^+}}   \
  \frac{1-\frac{g^2}{2x_k^+ x_j^-}}{1-\frac{g^2}{2x_k^-x_j^-}}
   \right)^{2i (u_k-u_j) }
    \ ,  \la{lei}\eea
    where the definitions of $x^\pm_k, u_k$ are the same as in \rf{ook}--\rf{yu}
    and $e^{ i \theta(p_j,p_k) } $ is an extra ``string'' S-matrix factor
    that distinguishes the string BA  from asymptotic gauge BA
    (indeed, after omitting this factor and setting $s=0$
    eq.\rf{ei} reduces to \rf{ook}).

In the $\su(1|1)$ sector we are interested in here the logarithm
of the  string Bethe equations  \rf{ei}   reads
 \bea &&p_k=2\pi\frac{n_k}{L}+\frac{1}{iL}\sum_{j\neq
k}^M \Big(
\log\frac{1-\frac{g^2}{2x_k^-x_j^+}}{1-\frac{g^2}{2x_k^+x_j^-}} +
2i (u_{k}-u_{j})
\log\frac{1-\frac{g^2}{2x_k^-x_j^+}}{1-\frac{g^2}{2x_k^+x_j^+}}
\frac{1-\frac{g^2}{2x_k^+x_j^-}}{1-\frac{g^2}{2x_k^-x_j^-}}\Big)
\, .\nonumber \\ \label{QSBE} \eea Here $n_k$ are the excitation
numbers
and $L=J+\sfrac{1}{2}M$.
As soon as
the momenta $p_k$ solving \rf{QSBE}
are found,  the energy can be computed by using the formula
(\ref{jy}).

\subsection{Weak coupling expansion}

While the  equations (\ref{QSBE}) were
originally found  by ``discretising'' the  integral equations which appear in
  the semiclassical string theory where $g$ is large \ci{kmmz},
 they also admit a regular { weak-coupling limit} $g\to 0$
\cite{bies}.
 Assuming the same distribution of the numbers $n_k$
as in the weak-coupling gauge theory \rf{kol}, one can solve equations
(\ref{QSBE}) perturbatively. In particular, few leading terms of
the momentum $p_k$ read \bea p_k =&& { 2 \pi n_k \ov L} - g^2 \sin
{ 2 \pi n_k \ov L} + { 1 \ov 4} g^4  ( 2
\sin { 2 \pi n_k \ov L} + 5 \sin { 4 \pi n_k \ov L}) \no \\
&&- \  { 1 \ov 12} g^6  ( -23 \sin { 2 \pi n_k \ov L} +64  \sin { 4
\pi n_k \ov L}  +  10  \sin { 6 \pi n_k \ov L}
    )  + {\cal O}(g^8)   \ , \  \la{ppQSBE} \eea
This leads to the following expansion for the energy (\ref{jy})
\bea \label{enQSBEweak} \frac{\Delta
}{L}&=&\frac{3}{2}+2g^2-4g^4+\frac{25}{2}g^6-
\frac{601}{12}g^8+\frac{2849}{12}g^{10} + ... \eea As expected,
the  first two  (one-loop $g^2$ and two-loop $g^4$) coefficients
are the same as in  \rf{eee} but the two series differ starting
with $g^6$. Note, however, that again the series is sign-alternate
and should have a finite radius of convergence.
Compared to the gauge Bethe ansatz equations  of the previous
section, here it seems even more challenging to try to find the
solution of the equations \rf{QSBE} in a closed form.

Let us  recall     that  the
equations (\ref{QSBE}) are known
 to
receive the $1/g \sim 1/\sqrt{\lambda}$
corrections \cite{szz,bt}  (required in order to reproduce
 quantum string results)
 which
could be universally incorporated \cite{bt} in the infinite set of
functions $c_r(\lambda)$.
These functions define a more general
{\it interpolating} string Bethe ansatz \cite{afs} and  should  lead to
a  modification of the weak coupling expansion (\ref{enQSBEweak})
which, hopefully, should   agree with that  found on  the gauge theory
side in the large $L$ limit.

\subsection{Strong  coupling expansion}

Since we  are interested in the highest-energy state  with $M=L$
impurities, the expression  for the energy \rf{jy}  suggests that
the  maximal energy would be attained if all $L$  momenta $p_k$
were equal  to $\pi$ (modulo $2 \pi n$.  Then all $  \sin {p_k \ov
2}=1$ and at strong coupling $E \to   { \sql \ov \pi}  L $.
However, all momenta must be distinct  since otherwise  the wave
function vanishes due to Fermi statistics of excitations, i.e.  one
is to choose some non-trivial distribution for $p_k$.
 As a result,
the coefficient $\frac{E}{\sql L}$   will be less than 1 (it was $
{3 \sqrt 3 \ov 2  \pi }\approx 0.83$   in the gauge Bethe ansatz
case \rf{qqw}).

We expect  the energy \rf{jy}  to scale as $\sql L$, so  we should assume
that the leading term in the strong-coupling expansion of momenta should be
constant, i.e. as in  \rf{gghh},
$p_k={p^{(0)}_k} +\frac{p^{(1)}_k}{\sql }+\ldots $,\foot{As was discussed in \cite{afs}, for short strings
a natural expansion of momenta is
$p_k=\frac{p^{(0)}_k}{\l^{1/4} }+\frac{p^{(1)}_k}{\l^{1/2} }+\ldots
$
leading to the  $\sqrt[4]{\lambda}$ scaling of the energy of the
corresponding states.} where  ${p^{(0)}_k}$ should be again
distributed, say on $(- \pi, \pi)$.

\medskip

Extracting this distribution from the  strong coupling expansion
of the string Bethe ansatz (\ref{QSBE}) appears to be  more subtle
than in the gauge Bethe ansatz case of section 2.3.
Observing that according to \rf{ooo}
\be \la{ukk}
u(p)_{_{\l \to \infty}} \  \to \  { \sql \ov 2\pi} \ep(p)  \cos {p\ov 2}   \ ,
\ee
 where   $\ep(p) $ is the sign factor defined in \rf{stro},
 we find that in the
limit  when $\l\to \infty$ the term proportional to $u_{k}-u_j $ in the
eq.(\ref{QSBE}) provides a dominant contribution
\bea \la{kopl}    {  i \sql  \ov \pi } \sum_{j\neq k}^M \Big[ \ep(p_k)
 \cos { p_k \ov 2} -\ep(p_j) \cos { p_j \ov
2}\Big] \log { 1- \ep(p_k) \ep(p_j)  \cos { p_k-p_j \ov 2} \ov
 1- \ep(p_k) \ep(p_j)  \cos { p_k+ p_j \ov 2}  }\, \ ,
 \eea
where  again for simplicity we
 renamed  $p_k^{(0)}\to p_k$.
 Setting  $M=L$  and assuming that there exists a solution
 with all $\ep(p_k) =1$
 we obtain the following  non-linear equations for the momentum distribution
\bea \label{vne}
 \sum_{j\neq k}^L \Big( \cos { p_k \ov 2} -
  \cos { p_j \ov 2}\Big) \log {
\sin^2 { p_k-p_j \ov 4} \ov \sin^2 { p_k+p_j \ov 4} } =0\, .\eea
It is unclear at  the moment how to find a solution of this set of
non-linear equations assuming that $p_k$
 obey an additional constraint
$\e(p_k)=1$. Moreover, it is also unclear if the resulting
expansion around this solution will be regular. Nevertheless, once
a solution to eq.(\ref{vne}) is found, the energy of this state is
guaranteed to have the same $\sqrt{\lambda}$ scaling behavior as
found in the strong-coupling gauge theory. We also note that
treating $p_k\equiv x_k$ as positions on $M$ particles on a circle
of length the $2\pi$, eqs.(\ref{vne}) can be thought of  as equations
determining an equilibrium configuration $\frac{\pa U }{\pa
x_k}=0$ for some potential $U$. It would be important to develop
this interpretation further, in particular, to see whether
solutions of eq.(\ref{vne}) can be related to zeros of some known
orthogonal polynomials.

It is interesting to note that the strong-coupling equations
\rf{kopl} appear to be universal: they  are found from \rf{ei} for
any value of the power $s$. Hence   exactly the  same
problem appears  in determining the maximal-energy  state as
described by the string Bethe ansatz also in the $\su(2)$ sector.
Moreover, the leading term in the strong-coupling expansion of the
energy will then be the same as in the $\su(1|1)$ and $\su(2)$
sectors.\foot{A similar  universality  is found in  the spectrum of short strings in the
strong-coupling (``flat-space'')  limit.}

\medskip

Let us mention  also the large $L$ form of the above Bethe equations.
In  the
  momentum representation of the string Bethe ansatz equations
  the analog of \rf{uir} becomes
\be 1=2\pi   \td \rho(p) + \frac{1}{i}\int  dq \ \td \rho(q)
\frac{\pa}{\pa p }  \td K(p,q)  \ , \la{kjv} \ee where $\td K(p,q)
\equiv  K(u(p),v(q))$ is found by using \rf{QSBE},\rf{yu}.
Taking the strong-coupling limit {\it assuming} that $p$ is fixed
in this limit, we get \be \la{ihh} \td K(p,q) =   \sql \td K_1
(p,q)  + \td K_2 (p,q)  \ , \ \ \ \ \ \ \ \ \ \ \ \ee where \be
 \td K_2  =  \log \frac{1- e^{  {i \ov 2} (p-q)}  \ep (p) \ep (q)  }
{1-    e^{ - {i \ov 2} (p-q) } \ep (p) \ep (q) }  = {i \ov 2}
(p-q)  + \const  \ ,
\ee  \be \la{ssa}
  \td K_1  =  {  i  \ov \pi}  [  \ep (p)  \cos { p \ov 2} - \ep (q)
\cos { q \ov 2}]
 \log\bigg[
\frac{1- e^{  {i \ov 2} (p-q)}  \ep (p) \ep (q)  } {1-    e^{ - {i
\ov 2} (p+q) } \ep (p) \ep (q) } \frac{1- e^{  -{i \ov 2} (p-q)}
\ep (p) \ep (q)  } {1-    e^{  {i \ov 2} (p+q) } \ep (p) \ep (q) }
\bigg]   \ . \ee
If as in the gauge BA case
 the leading term in $\td \rho(p)$ for the highest-energy state
does not depend on $\lambda$, we need  to ensure that the leading
term $\td K_1$  does not contribute  to \rf{kjv}.
Assuming  that for
our solution  $\td \rho$=const we are to  satisfy \be \la{yye}
\int  dq \  \frac{\pa}{\pa p }  \td K_1 (p,q) =0 \ee
which is the continuum analog of the vanishing of \rf{kopl}.
The same   discussion of the leading large $\lambda$ asymptotics
 applies also to the $\su(2)$ case.
It is not clear how to satisfy the condition  \rf{yye}, and this may be
indicating a potential problem in direct application of the AFS-type Bethe
ansatz to determining the highest-energy state at strong coupling.

\subsection{Comments on the spectrum
of reduced  $\su(1|1)$ string model}

In the absence of direct information about
the structure of  exact string spectrum  one may try  to draw some
lessons from ``reduced'' models obtained by truncating the string degrees
of freedom  at the classical level and then quantizing the remaining modes.
While the  truncation and quantization procedures  are not expected to commute,
the spectrum of reduced model may still reflect certain features of the exact
string spectrum.
Let us finish this section with a review of the structure of the spectrum of the
reduced $\su(1|1)$ model \cite{Alday:2005jm,af2}.

At the classical level the $\su(1|1)$ sector of the $\AdS$ string
theory can be defined as a consistent truncation of the
superstring equations of motion  \cite{Alday:2005jm}. In the
light-cone gauge this model reduces to the theory of a free
massive world-sheet Dirac fermion,  and therefore, can be easily
quantized \cite{af2}.
 The corresponding spectrum correctly
reproduces the leading $1/J$-corrections to the energy of the
plane-wave states but at higher orders leads to the  results which
are different from the ones predicted by the ``string'' Bethe
equations (\ref{QSBE}).
\foot{Truncating the classical superstring
equations of motion in the {\it  temporal}  gauge one finds a new non-linear
classically-integrable fermionic model \cite{Alday:2005jm}.
 This non-linear AAF model, however, is not power-counting renormalizable at
the quantum level and is not readily solvable
(though its low-energy 2-body S-matrix and thus the corresponding Bethe ansatz
 may be computed  assuming quantum integrability \ci{kloz}).
 While at the classical level the light-cone gauge and temporal gauge reduced models are
 equivalent, this is not so at the quantum level.
The difference  does not appear  at order $1/J$  (the near plane-wave limit)
but the two models start to disagree at order  $1/J^2 $
where the AAF model first gets nontrivial UV
divergencies and  where the omitted interactions
with other sectors (which cancel the divergences in the full superstring theory)
 become important. That  means,  in particular,
 that from the string theory perspective
 one cannot trust the non-linear AAF model more than the free light-cone gauge model
 beyond the $1/J$ level.}
 Assuming that the AFS ansatz does in fact
represent the correct quantum string spectrum, one natural
interpretation of this
 difference is  that beyond the leading order the
string modes which were truncated away at the classical level
start to contribute.
Given that  the quantum version of the reduced
$\su(1|1)$ model leads to an approximate  description of the
AFS-type  Bethe equations in the $\su(1|1)$ sector in
the region  of small momenta $p_k$
 it is instructive to review
the scaling behavior of the energy of different states in this model.

The momenta $p_k$ of the elementary excitations in the reduced
$\su(1|1)$ model are subject to the following Bethe-type equations
 \cite{af2} (see also \cite{stau}) \bea \label{honest}
Jp_k=2\pi n_k+\frac{1}{2} \sum_{j\neq
k}^M\Big(p_j\sqrt{1+\frac{\lambda
p_k^2}{4\pi^2}}-p_k\sqrt{1+\frac{\lambda p_j^2}{4\pi^2}}\Big)\,
, \eea where the expression under the sum  is the logarithm of
the two-body ``string S-matrix'' and $n_k$ are integers. The energy of
the corresponding state is \bea \label{enhonest}
E=J+\sum_{j=1}^M\sqrt{1+\frac{\lambda p_j^2}{4\pi^2}}\, . \eea The
spectrum contains two types of string excitations: short strings
with vanishing winding $m=\sum_{k=1}^M p_k$ and long strings with
$m\neq 0$. For short strings the energy scales in the large
$\lambda$ limit as $\sqrt[4]{\lambda}$ and this scaling is
perfectly consistent with one predicted by the AFS Bethe   equations.

For long strings the situation is different.  Summing up
eqs.(\ref{honest}) we get a condition \bea \label{wfen}
J\sum_{k=1}^M p_k=2\pi \sum_{k=1}^M n_k ~~~~\Longrightarrow
~~~~Jm=\sum_{k=1}^M n_k\, . \eea Thus, for non-vanishing winding
we have to assume that the momenta have  the following
expansion\footnote{Here we assume that all $p_k^{(0)}$ are
non-zero. One can consider a possibility that only a part of
$p_k^{(0)}$ is non-zero. This will lead to modifications of the
expansions discussed below.} \bea \label{mexp}
p_k=p_k^{(0)}+\frac{p_k^{(1)}}{\sqrt{\lambda}}+...\, ,
\label{sce}\eea where the leading term $p_k^{(0)}$ is constant.
Then in large $\lambda$ limit only the second ``string S-matrix"
term in \rf{honest} matters. Expanding eq.(\ref{honest}) we find
at the two leading orders the following equations \bea \la{onj}
\sqrt{\lambda}:~~~~&&
\sum_{j\neq k}(p_j^{(0)}|p_k^{(0)}|-p_k^{(0)}|p_j^{(0)}|)=0 \, ,\\
\nonumber
 \lambda^0:~~~~&& Jp_k^{(0)}-2\pi n_k
-\frac{1}{4\pi}\sum_{j\neq k}
\Big(p_k^{(0)}p_j^{(1)}+p_j^{(0)}p_k^{(1)}\Big) \Big({\rm
sign}(p_k^{(0)})- {\rm sign}(p_j^{(0)}) \Big)=0 \, .\eea The first
equation implies
$$
\sum_{j=1}^M|p_j^{(0)}|=m\frac{|p_k|^{(0)}}{p_k^{(0)}}=m~{\rm
sign}p_k^{(0)} ~~~~{\rm for~any~}k.
$$
Thus $p_k^{(0)}$ must be either all positive or all negative.
Assuming that they are all positive we conclude  from the second
equation that $p_k^{(0)}=\frac{2\pi n_k}{J}$, where all $n_k>0$.
It is rather interesting that the leading equation (\ref{onj}) is
satisfied identically provided $p_k^{(0)}$ are all positive or all
negative. One can go further and find that with our assumption of
positivity of $n_k$ the next order in the expansion of the Bethe
equations (\ref{honest}) leads to the determination of
$p_k^{(1)}$: \bea \frac{1}{\sqrt{\lambda}}:~~\ \ \ \ \
p_k^{(1)}=\frac{\pi}{2J}\sum_{k\neq
j}^M\Big(\frac{n_j}{n_k}-\frac{n_k}{n_j}\Big)\, . \eea Thus, the
strong coupling expansion of a long string configuration is
well-defined and the leading momenta are \be p_k=\frac{2\pi n_k
}{J}+\frac{\pi}{2J\sqrt{\l}}\sum_{k\neq
j}^M\Big(\frac{n_j}{n_k}-\frac{n_k}{n_j}\Big)\,+\ldots \ . \ee The
expansion for the energy is therefore \bea
E=\frac{\sqrt{\lambda}}{J }\sum_{k=1}^M
n_k+J+\frac{J}{2\sqrt{\lambda}}\sum_{k=1}^M
\frac{1}{n_k}-\frac{J}{16\lambda}\sum_{k,j=1}^M\frac{(n_k^2-n_j^2)^2}{n_k^3
n_j^3}+\ldots \, . \eea Thus in the case of non-trivial winding
(long strings) the energy scales as $\sqrt{\lambda}$ \cite{af2}.
Here the  leading term in the
 energy scales as $ E = {\sqrt{\l}\ov J}  \sum_k^M n_k,$ i.e.
 as $ E= \sql m $   if we use \rf{wfen}.\foot{The $\sql m$ behavior of
 energy of long wound strings was observed earlier
 in more general context in \ci{arfr}.}
 However,  the energy is not bounded from above: the string
excitation numbers $n_k$  can be arbitrarily large
 making the winding and energy unbounded.
 One might  expect that true quantum string states
 will develop periodicity in $n_k$ so that $m$ will be bounded
 from above by a number of order $J$, and thus
 there will exist a maximal-energy state.

Indeed, one may speculate  that the non-trivial dependence
of the reduced model energy on  the winding number $m$
in this  simplified model is an artifact of the perturbative
expansion while in the full quantum string
theory the explicit dependence on $m$ will
be traded for a periodic dependence on $n_k$.
Eq.(\ref{wfen}) shows that the winding is not an
independent variable but can be expressed in terms of the
excitation numbers $n_k$. If  the exact dispersion relation
of the quantum string theory
 will indeed  appear to be
periodic, i.e.  invariant under the shift
$p_k\to p_k+2\pi $ ($n_k\to n_k+J$)
then one will  always be able to choose
a momentum distribution which  has zero winding without changing the
value of the
energy which  will then  be bounded from above
due to the compactness of the phase space.

In spite of the fact that the reduced model  does not appear to
describe exact quantum string states  it exhibits the following
features which we  expect to find in  the genuine quantum string theory:
(i) it suggests that the energy can indeed have
$\sqrt{\lambda}$ behavior at large $\lambda$, and
(ii)
to get the $\sqrt{\lambda}$ scaling of the energy, the
momenta of the corresponding elementary excitations should have the same
 large $\lambda$ expansion  as in eq.(\ref{mexp}).
These were the properties which were implemented in the solution
of the gauge Bethe ansatz equations in section 2  and also the
ones we  were assuming above in trying to solve  the AFS-type
string Bethe equations.  In fact, the expansion of \rf{honest}
 is very similar to the expansion of the string Bethe equations \rf{QSBE} in the strong coupling
limit (cf. \rf{onj} and \rf{kopl},\rf{vne}).


\section{Summary}

The  maximal energy state in the $\su(1|1) $  sector
 we discussed in this paper is special:
 this is one of very few cases when
 we know  explicitly  the exact quantum  operator as well the corresponding
  distribution of  momenta   describing
  it as a solution of  the gauge/string  Bethe ansatz equations.

Let us  summarize the asymptotic expansions
for the momentum distributions we have found above (see Table 1).

{\footnotesize
\bea \nonumber
\begin{array}{c|c|c}
            & {\rm Gauge~~BA} & {\rm String~~BA} \\
            \hline
                  &                          & \\
{\rm Weak}        &  p_k=\frac{2\pi k}{L}+{\cal O}(\l)   & p_k=\frac{2\pi k}{L}+{\cal O}(\l)\\
{\rm coupling}    &     & \\
\hline
 &  &  \\
 {\rm Strong}      &  ~~~p_k=\frac{4\pi k}{3L} + {\pi \ov 3}
( 1 + { 1 \ov L})+{\cal O}\Big(\frac{1}{\sqrt{\lambda}}\Big)~~
   & ~~~p_k=p^{(0)}_k(L)   +{\cal O}\Big(\frac{1}{\sqrt{\lambda}}\Big) \\
{\rm coupling}    &     &
\end{array}
\eea }\begin{center} Table 1. Leading momentum behavior of
maximal-energy solution of gauge \\\noindent \hspace{-2cm} and
string  Bethe ansatze in the $\su(1|1)$ sector.
\end{center}

We have argued that the same  qualitative features of the maximal
energy state as found from  the gauge theory BDS-type gauge Bethe
ansatz   should  appear also for the associated  quantum string
state but we were unable to solve the strong-coupling limit of the
conjectured AFS-type string Bethe equations explicitly. In fact,
the strong-coupling behavior of the corresponding integral
equation kernels in these  ``gauge'' and ``string'' Bethe ansatze
 appears to be very different. This may be seen as another
indication of a need for clarifying  the structure of the
``string'' Bethe ansatz and, in particular, for understanding
whether and how it captures the higher-energy  tail of the string
spectrum.

\medskip


\newpage
{\bf Note added}:
While this paper was prepared  for publication
we learned  about a very  closely related  unpublished work  by D.
Serban and M. Staudacher \ci{serst} who also found the
perturbative solutions of the asymptotic  gauge and string  Bethe
ansatze  in the $\su(1|1)$ sector for the purely fermionic state
(in particular, the expressions in eqs. \rf{eee}, \rf{sss}, \rf{enQSBEweak}).

\bigskip

\section*{Acknowledgments }

We are  grateful to N. Beisert, S. Frolov,  R. Roiban, D. Serban, M. Staudacher
 and  K. Zarembo for
very useful
 discussions and comments.
  The work of
G.~A. was supported in part by the European Commission RTN
programme HPRN-CT-2000-00131, by RFBI grant N05-01-00758, by NWO
grant 047017015 and by the INTAS contract 03-51-6346.
 The work of A.T. was supported  in part by the
 PPARC grant PPA/G/O/2002/00474,
  the INTAS grant  03-51-6346, the DOE grant DE-FG02-91ER40690
 and the RS Wolfson award.



\begin{thebibliography}{20}
\bibitem{mz1}
J.~A.~Minahan and K.~Zarembo, ``The Bethe-ansatz for N = 4 super
Yang-Mills,'' JHEP {\bf 0303}, 013 (2003), hep-th/0212208.

\bi{bs} N. Beisert and M. Staudacher, ``The N=4 SYM integrable
super spin chain'', Nucl.\ Phys.\ B {\bf 670}, 439 (2003),
hep-th/0307042.

\bibitem{Bena:2003wd}
  I.~Bena, J.~Polchinski and R.~Roiban,
  ``Hidden symmetries of the $\AdS$ superstring,''
  Phys.\ Rev.\ D {\bf 69}, 046002 (2004), hep-th/0305116.

\bibitem{Arutyunov:2003uj}
  G.~Arutyunov, S.~Frolov, J.~Russo and A.~A.~Tseytlin,
  ``Spinning strings in $\AdS$ and integrable systems,''
  Nucl.\ Phys.\ B {\bf 671} (2003) 3, hep-th/0307191.



\bibitem{BMN}
  D.~Berenstein, J.~M.~Maldacena and H.~Nastase,
  ``Strings in flat space and pp waves from N = 4 super Yang Mills,''
  JHEP {\bf 0204} (2002) 013, hep-th/0202021.

\bi{ft}
  S.~Frolov and A.~A.~Tseytlin,
  ``Semiclassical quantization of rotating superstring in $\AdS$,''
  JHEP {\bf 0206}, 007 (2002), hep-th/0204226.
``Multi-spin string solutions in $\AdS$,''
  Nucl.\ Phys.\ B {\bf 668}, 77 (2003), hep-th/0304255.
  N.~Beisert, J.~A.~Minahan, M.~Staudacher and K.~Zarembo,
  ``Stringing spins and spinning strings,''
  JHEP {\bf 0309}, 010 (2003), hep-th/0306139.
N.~Beisert, S.~Frolov, M.~Staudacher and A.~A.~Tseytlin,
  ``Precision spectroscopy of AdS/CFT,''
  JHEP {\bf 0310}, 037 (2003), hep-th/0308117.


\bi{kmmz} V.~A.~Kazakov, A.~Marshakov, J.~A.~Minahan and
K.~Zarembo,
  ``Classical / quantum integrability in AdS/CFT,''
  JHEP {\bf 0405}, 024 (2004), hep-th/0402207.


\bi{lip}
 A.~V.~Kotikov, L.~N.~Lipatov, A.~I.~Onishchenko and V.~N.~Velizhanin,
  ``Three-loop universal anomalous dimension of the Wilson operators in N =  4
  SUSY Yang-Mills model,''
  Phys.\ Lett.\ B {\bf 595}, 521 (2004)
  [Erratum-ibid.\ B {\bf 632}, 754 (2006)], hep-th/0404092.
A.~V.~Kotikov, L.~N.~Lipatov and V.~N.~Velizhanin,
  ``Anomalous dimensions of Wilson operators in N = 4 SYM theory,''
  Phys.\ Lett.\ B {\bf 557}, 114 (2003), hep-ph/0301021.

\bi{ryz} A.~V.~Ryzhov and A.~A.~Tseytlin,
  ``Towards the exact dilatation operator of N = 4 super Yang-Mills theory,''
  Nucl.\ Phys.\ B {\bf 698}, 132 (2004), hep-th/0404215.


\bibitem{rss}
  A.~Rej, D.~Serban and M.~Staudacher,
  ``Planar N = 4 gauge theory and the Hubbard model,''
  hep-th/0512077.

\bi{zar}
  K.~Zarembo,
  ``Antiferromagnetic operators in N = 4 supersymmetric Yang-Mills theory,''
  Phys.\ Lett.\ B {\bf 634} (2006) 552, hep-th/0512079.


\bibitem{bds}
  N.~Beisert, V.~Dippel and M.~Staudacher,
  ``A novel long range spin chain and planar N = 4 super Yang-Mills,''
  JHEP {\bf 0407}, 075 (2004), hep-th/0405001.


\bibitem{afs}
  G.~Arutyunov, S.~Frolov and M.~Staudacher,
  ``Bethe ansatz for quantum strings,''
  JHEP {\bf 0410}, 016 (2004), hep-th/0406256.

\bibitem{rtt}
  R.~Roiban, A.~Tirziu and A.~A.~Tseytlin,
  ``Slow-string limit and 'antiferromagnetic' state in AdS/CFT,''
  hep-th/0601074.

\bibitem{BEI}
  N.~Beisert,
  ``The complete one-loop dilatation operator of N = 4 super Yang-Mills
  theory,''
  Nucl.\ Phys.\ B {\bf 676}, 3 (2004), hep-th/0307015.


\bibitem{stau}
  M.~Staudacher,
  ``The factorized S-matrix of CFT/AdS,''
  JHEP {\bf 0505}, 054 (2005), hep-th/0412188.


\bibitem{beist}
  N.~Beisert and M.~Staudacher,
  ``Long-range PSU(2,2$|$4) Bethe ansaetze for gauge theory and strings,''
  Nucl.\ Phys.\ B {\bf 727}, 1 (2005), hep-th/0504190.

\bibitem{bt}
  N.~Beisert and A.~A.~Tseytlin,
  ``On quantum corrections to spinning strings and Bethe equations,''
  Phys.\ Lett.\ B {\bf 629} (2005) 102, hep-th/0509084.



\bi{szz} S.~Schafer-Nameki, M.~Zamaklar and K.~Zarembo,
  ``Quantum corrections to spinning strings in $\AdS$ and Bethe ansatz:
  A comparative study,''
  JHEP {\bf 0509}, 051 (2005), hep-th/0507189;
  S.~Schafer-Nameki and M.~Zamaklar,
  ``Stringy sums and corrections to the quantum string Bethe ansatz,''
  JHEP {\bf 0510} (2005) 044, hep-th/0509096;
  S.~Schafer-Nameki,
  ``Exact expressions for quantum corrections to spinning strings,''
  hep-th/0602214.



\bibitem{Alday:2005jm}
  L.~F.~Alday, G.~Arutyunov and S.~Frolov,
  ``New integrable system of 2dim fermions from strings on $\AdS$,''
  JHEP {\bf 0601} (2006) 078, hep-th/0508140.

\bi{af2} 
  G.~Arutyunov and S.~Frolov,
  ``Uniform light-cone gauge for strings in $\AdS$: Solving su(1$|$1)
  sector,''
  JHEP {\bf 0601} (2006) 055, hep-th/0510208.



 \bi{beis}
N.~Beisert, ``The dilatation operator of N = 4 super Yang-Mills
theory and integrability,'' Phys.\ Rept.\  {\bf 405}, 1 (2005),
hep-th/0407277.

\bi{beii} N.~Beisert,
  ``The su(2$|$3) dynamic spin chain,''
  Nucl.\ Phys.\ B {\bf 682}, 487 (2004), hep-th/0310252.
C.~G.~Callan, J.~Heckman, T.~McLoughlin and I.~J.~Swanson,
  ``Lattice super Yang-Mills: A virial approach to operator dimensions,''
  Nucl.\ Phys.\ B {\bf 701}, 180 (2004), hep-th/0407096.



\bi{hult} L. Hulthen, Ark.f.Mat.Astr.Fys. {\bf 26A:11} (1938) 1.


\bi{bies}
 N.~Beisert,
  ``Spin chain for quantum strings,''
  Fortsch.\ Phys.\  {\bf 53}, 852 (2005), hep-th/0409054.

\bibitem{kloz}
  T.~Klose and K.~Zarembo,
  ``Bethe ansatz in stringy sigma models,''
  hep-th/0603039.

\bibitem{arfr}
  G.~Arutyunov and S.~Frolov,
  ``Integrable Hamiltonian for classical strings on $\AdS$,''
  JHEP {\bf 0502} (2005) 059, hep-th/0411089.

\bi{serst}
D. Serban and M. Staudacher, unpublished  draft (2005).




\end{thebibliography}
\end{document}